\documentclass[prl,twocolumn,preprintnumbers,amsmath,amssymb,apsrev]{revtex4}

\usepackage{graphicx}
\usepackage{dcolumn}
\usepackage{bm}
\usepackage{subfigure}
\usepackage[sort&compress]{natbib}

\begin{document}

\title{STM and \emph{ab initio} study of holmium nanowires on a Ge(111) Surface}
\author{C. Eames, C. Bonet, M. I. J. Probert and S. P. Tear}\email[Corresponding author. E-mail:\,\,\,]{spt1@york.ac.uk}
\affiliation{Department of Physics, University of York, York YO10
5DD, United Kingdom}
\author{E. W. Perkins}
\affiliation{School of Physics and Astronomy, University of
Nottingham, Nottingham NG7 2RD, United Kingdom}

\date{November 2006}

\begin{abstract}
A nanorod structure has been observed on the Ho/Ge(111) surface
using scanning tunneling microscopy (STM). The rods do not require
patterning of the surface or defects such as step edges in order
to grow as is the case for nanorods on Si(111). At low holmium
coverage the nanorods exist as isolated nanostructures while at
high coverage they form a periodic 5$\times$1 structure. We
propose a structural model for the 5$\times$1 unit cell and show
using an \emph{ab initio} calculation that the STM profile of our
model structure compares favourably to that obtained
experimentally for both filled and empty states sampling. The
calculated local density of states shows that the nanorod is
metallic in character.
\end{abstract}

\maketitle

Self-assembled nanorods on surfaces have potential applications in
optoelectronic and microelectronic devices and some interesting
properties have already been reported for such structures
\cite{Yeom, Segovia, Losio}.

On the Si(001) surface self-assembled nanorods have been observed
for many adsorbates including the group III-IV metals (Ref.
\cite{Wang} and references therein) and the rare-earth (RE) metals
(\,\cite{Liu, Ragan, Kuzmin, Chen} and references therein). The
Si(111) surface has threefold symmetry and is not as favourable a
host to nanorod growth. Gd and Pb nanorods have been grown in the
lee of step edges along the [110] direction \cite{Kirakosian,
Hoque}. On the terraced regions the way forward is to deposit an
adsorbate on a prestructured adsorbate/Si(111) surface. In this
way Pb nanorods have recently been grown on the Sm/Si(111) surface
\cite{Palmino}.

Germanium is attracting renewed interest as a semiconductor
because it possesses a high hole carrier mobility. On Ge(001) no
nanowires involving RE metals have been reported. On Ge(111) a one
monolayer (ML) deposit of a RE metal forms a two-dimensional
1$\times$1 structure \cite{Spence} after annealing at
$\approx500^{\,\circ}$C. A similar structure is formed in
Si(111)1$\times$1 RE systems (Ref. \cite{Bonet} and therein).
Pelletier \textit{et al.} briefly noted the formation of
metastable rod like structures on the Er/Ge(111)-$c$(2$\times$8)
coexisting with the 1$\times$1 structure \cite{Pelletier}.

In this paper we report the formation of holmium nanorods on the
Ge(111) surface which we have observed in a STM experiment and
whose structure and properties we have investigated using an
\emph{ab initio} density functional theory calculation. We have
deposited low coverages (0.1-0.15 ML) of Ho onto a clean substrate
held at $250^{\circ}$C and instead of the 1$\times$1 structure we
observe a series of isolated holmium nanorods. These are true
isolated nanostructures because they are not part of a periodic
reconstruction or rectangular islands. The nanorods have a
constant width that is very narrow compared to other nanorod
structures and they do not require step edges or patterning in
order to form. When the experiment is repeated using a higher Ho
coverage the nanorods exist as part of a periodic 5$\times$1
structure. We have performed a medium energy ion scattering (MEIS)
study and used this in conjunction with the STM data to
parameterise the structure to the extent that we can suggest a
quantitative model for the 5$\times$1 unit cell. The simulated STM
for the model structure is qualitatively compared to the STM
images obtained in the laboratory and the comparisons are
favourable.

The STM experiments were done with an Omicron Nanotechnology GmbH
microscope at a typical UHV base pressure of $\leq2 \times
10^{-10}$ mbar. The germanium substrate was cleaned by argon ion
bombardment followed by annealing at $\approx500^{\,\circ}$C for
about 15 minutes and \emph{in situ} low-energy electron
diffraction was used to check that a clean Ge(111)-$c$(2$\times$8)
surface had been made.

The sample was prepared by depositing 0.1 ML of Ho from a quartz
crystal calibrated evaporation source onto a clean
Ge(111)-$c$(2$\times$8) surface held at a temperature of $\approx
250\,^{\circ}$C that was monitored using a $k$-type thermocouple
attached to the sample stage. Figure 1 shows a STM image of a
large area on the surface. Isolated and well defined nanorods have
formed on flat regions clear of step edges on a surface that was
not patterned or pre-structured in any way. The rods typically
extend for $\approx 40$\,nm and they have a constant width. The
structures have been reproduced in several experiments.

\begin{figure}
  \includegraphics[width=3.5in]{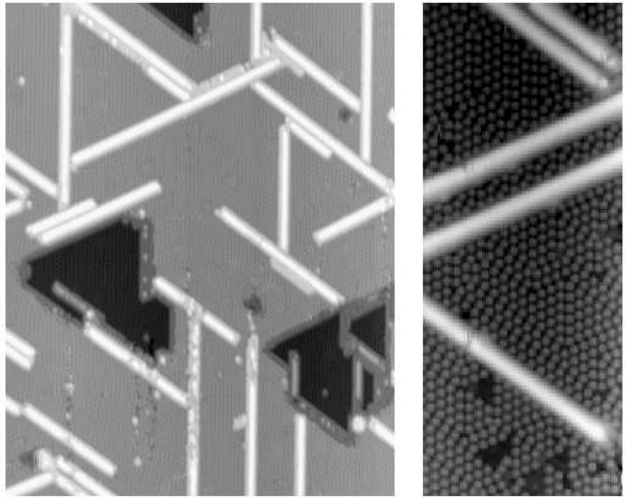}\\
  \caption{Two overview STM images of the nanorods formed with a low (0.1 ML) coverage of Ho.
  (Left) Large area (59 $\times$ 75 nm$^{2}$) filled states image taken with a sample bias of $-$\,2.0\,V and a tunneling
  current of 2\,nA. (Right) 14 $\times$ 33 nm$^{2}$ empty states image taken with sample bias +\,2.0\,V and a
  tunneling current of 2\,nA showing the surrounding clean Ge(111)-$c$(2$\times$8) with some domains of (2$\times$2)
  and $c$(4$\times$2) that are typically present on this substrate.}\label{fig1}
\end{figure}

\begin{figure}[t]
\centering \vspace{0.3cm}\subfigure[\,Filled states experiment.] {
    \label{fig:sub2:a}
    \includegraphics[width=2.0in]{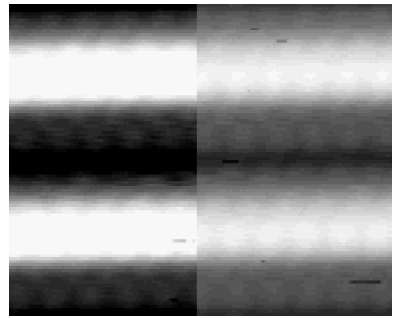}
} \hspace{1cm} \vspace{0.4cm} \subfigure[\,Filled states theory.] % caption for subfigure b
{
    \label{fig:sub2:b}
    \includegraphics[width=2.0in]{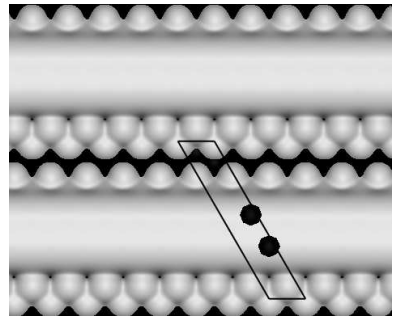}
} \caption{Measured (a) and simulated (b) filled states STM images
for the Ge(111)5$\times$1-Ho system. The tunneling current in the
experiment was 2\,nA. Both images correspond to a sample bias of
$-$\,2.0\,V and the image dimensions are 4.2 $\times$ 3.3
nm$^{2}$.}
\label{fig:sub2} % caption for the whole figure
\end{figure}

At a higher Ho coverage of 0.25 ML the rods stack in close
parallel proximity, forming small islands comprised of a periodic
5$\times$1 structure of which the isolated nanorods are a
precursor. The dimensions of the 5$\times$1 unit cell were
measured using surrounding areas of Ge(111)-$c$(2$\times$8) for
internal calibration. Figure \ref{fig:sub2:a} shows a filled
states image of a region 4.2 $\times$ 3.3 nm$^{2}$ in size that
contains two nanorods separated by a small gap taken from a region
that comprises seven nanorods side by side. The atoms within each
nanorod appear to occupy two distinct levels. With respect to the
Ge rest atom of the surrounding $c$(2$\times$8) reconstruction the
height of the lower level is 2.1 \AA\, and that of the nanorod
peak is 3.9 \AA. The width of the lower layer is 1.7 nm.

\begin{figure}[t]
\centering \subfigure[\,Empty states experiment.] {
    \label{fig:sub3:a}
    \includegraphics[width=2.0in]{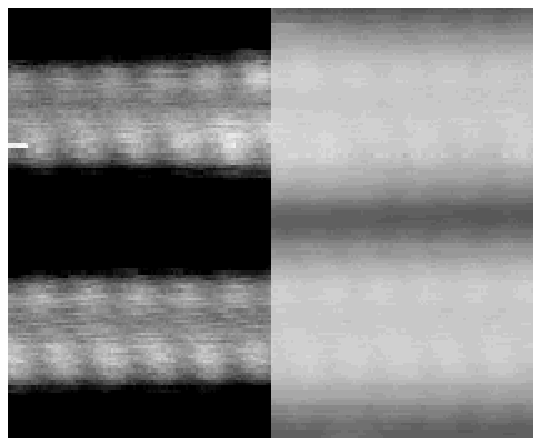}
} \hspace{1cm} \subfigure[\,Empty states theory.] % caption for subfigure b
{
    \label{fig:sub3:b}
    \includegraphics[width=2.0in]{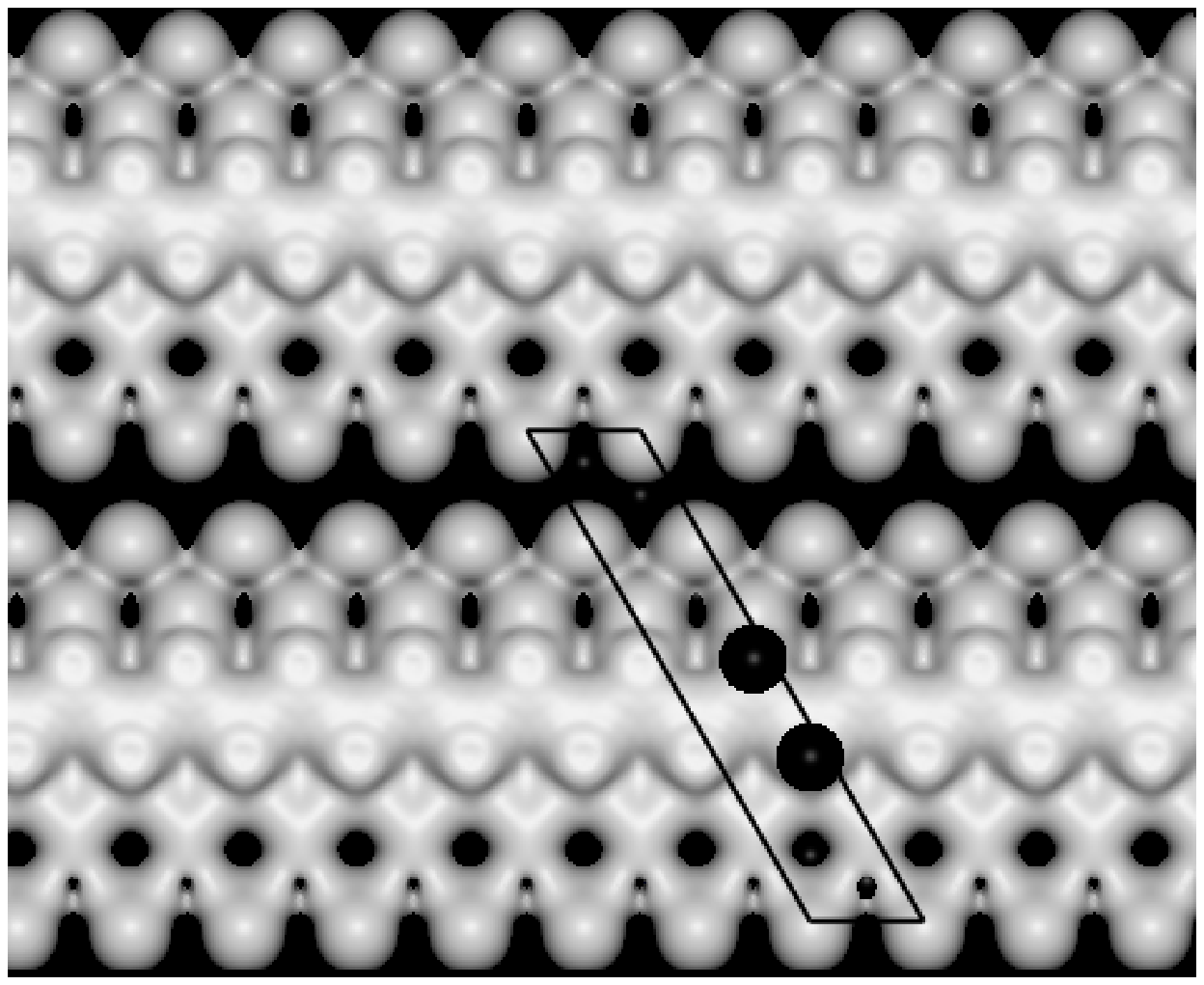}
} \caption{Measured (a) and simulated (b) empty states STM images
for the Ge(111) 5$\times$1 Ho system. The tunneling current in the
experiment was 2\,nA. Both images correspond to a sample bias of
+\,1.50\,V and the image dimensions are 4.2 $\times$ 3.3 nm$^{2}$.
The left half of (a) has been contrast adjusted so that the Ho
atoms within the top layer of the nanorod can be more clearly
discerned.}
\label{fig:sub3} % caption for the whole figure
\end{figure}

Figure \ref{fig:sub3:a} shows an empty states STM image of a
region 4.2 $\times$ 3.3 nm$^{2}$ in extent. Features within the
top layer of the nanorod can now be resolved, especially when the
image contrast is adjusted as in the left half of the figure. In
the empty states STM images the top layer of the nanorod was
measured as being 3.1 \AA \, above the Ge adatom layer (and thus
3.8 \AA \, above the rest atom layer since the adatom layer is
known to be 0.66 \AA \ above the rest atom layer \cite{Silfhout}).

Whilst we cannot claim that our layer height measurements
correspond to pure topography with no contribution from electronic
effects, the consistency of the filled and empty states
measurements and the large height difference of 3.8-3.9 \AA \,
from the nanorod peak to the Ge rest atoms allow us to conclude
that any model of the nanorod should involve a two-layer
structure.

A MEIS experiment in which incident H$^{+}$ ions were strongly
scattered in all directions by the Ho atoms was carried out at the
CCLRC Daresbury UK MEIS facility. The ion flux scattered from Ho
atoms did not show any dips at any emergent angle which might
indicate blocking by Ge atoms in a layer above the Ho atoms. We
thus interpret the two "bright" features per 5$\times$1 unit cell
in the empty states STM images as being associated with holmium
atoms forming the upper layer of the nanorod.

The Ge atoms in the lower level of the nanorod, 2.1~\AA \,above
the rest atoms of the $c$(2$\times$8) reconstruction, appear to be
too low to be a full Ge bilayer (which would have a 3.27 \AA \,
step height \cite{Silfhout}) and a STM image (not shown) in which
nanorods can be seen on adjacent terraces separated by a
monoatomic step supports this conclusion. Consideration of the
bonding requirements of the two trivalent Ho atoms per 5$\times$1
unit cell also leads to the conclusion that Ho atoms cannot be
adsorbed atop a simple bulk-terminated Ge surface since in such a
situation the Ge surface would provide only 5 dangling bonds to be
quenched by the two trivalent Ho atoms.

On the other hand, the lower layer of the nanorod is too high
above the rest atoms of the Ge substrate to be a simple adatom
layer (0.66 \AA \, above the rest atoms in clean
Ge(111)-$c$(2$\times$8) \cite{Silfhout}). A visual comparison with
surrounding areas of the $c$(2$\times$8) reconstruction in the
empty states STM images suggests an atomic density in the lower
nanorod layer which is greater than that of a dilute (e.g.
2$\times$2) adatom layer. Further information was obtained from
our STM observations of occasional faulted nanowire growth in
which the topmost Ho layer was sometimes absent from the nanorod
over a small region. Under these conditions the Ge layer was found
to be continuous across the width of the nanorod, extending across
the area that would normally be covered by the upper Ho layer. The
lower level of the nanorod was therefore identified as consisting
of a single layer of additional Ge atoms (with density around one
monolayer) atop the first bulk like Ge bilayer.

Given these considerations we propose the structure that is shown
in figure 4 in which there is a Ho nanorod atop an almost flat Ge
layer atop a bulk like Ge substrate. The structural parameters of
the 5$\times$1 supercell are available upon request.

\begin{figure}[h]
\centering \subfigure[\,Top view.] {
    \label{fig:sub4:a}
    \includegraphics[width=3.0in]{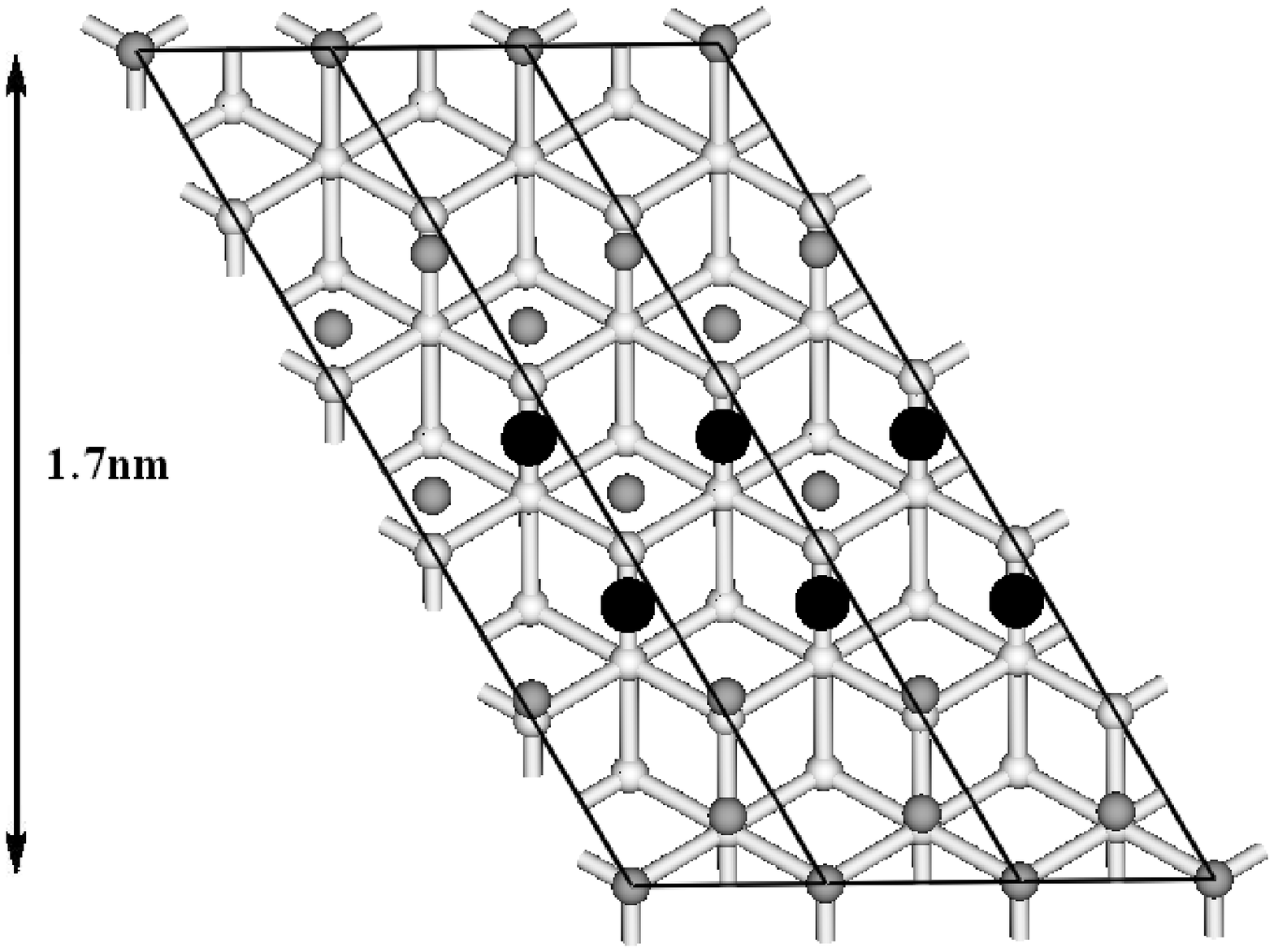}
} \hspace{1cm} \subfigure[\,Side view.] % caption for subfigure b
{
    \label{fig:sub4:b}
    \includegraphics[width=3.0in]{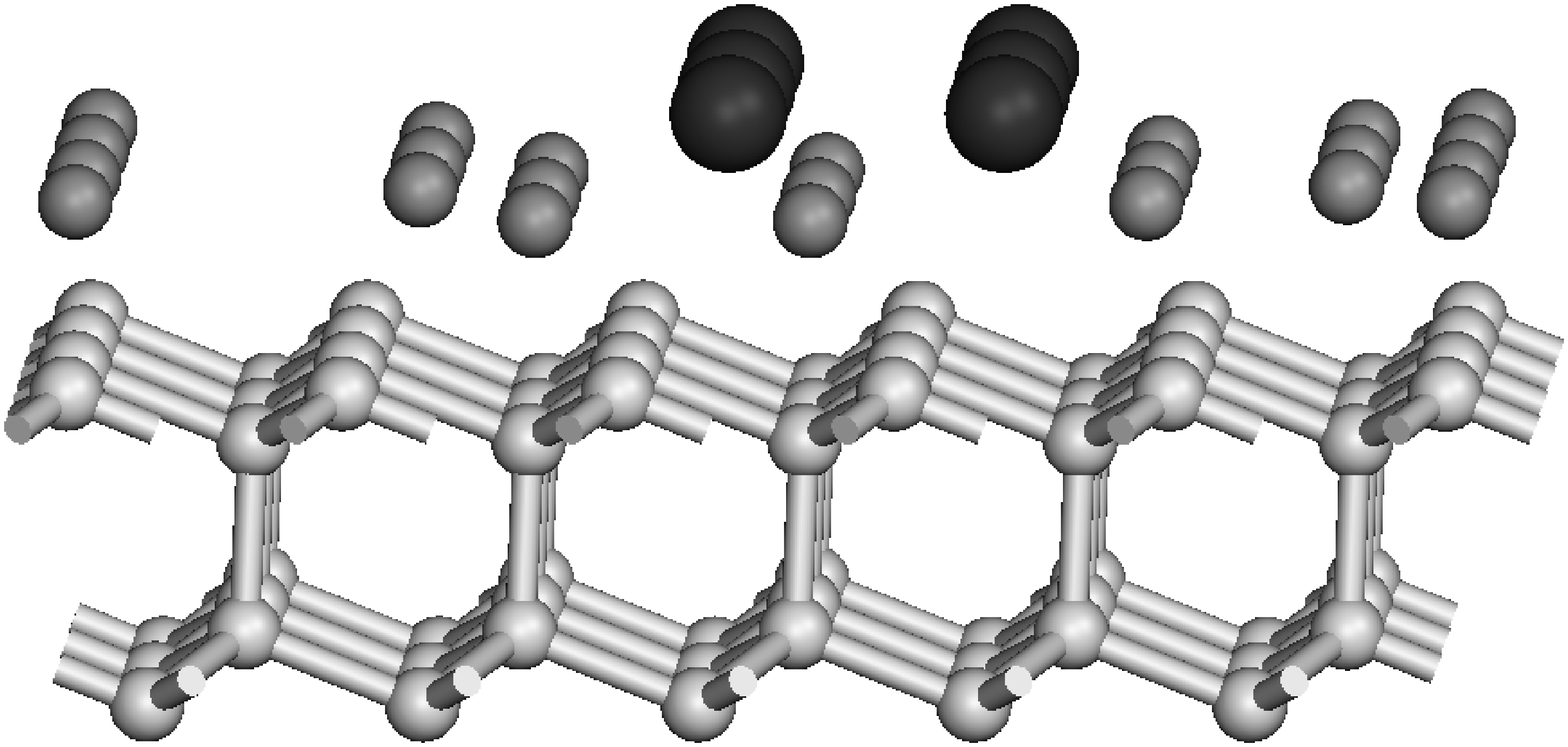}
} \caption{Two views of the Ge(111)5$\times$1-Ho system; (a) top
view in which the 5$\times$1 unit cell is outlined in black (b)
side view. Large black atoms are Ho, dark grey is reconstructed Ge
and light grey is bulk like Ge. }
\label{fig:sub4} % caption for the whole figure
\end{figure}

To validate the structural model we have calculated the STM image
that it would be expected to produce using the CASTEP \emph{ab
initio} density functional theory based code \cite{Segall}. The
generalised gradient approximation was used to model exchange and
correlation effects. The electronic wave function was expanded in
a plane wave basis set with a cutoff energy of 360 eV. The ionic
cores were represented with ultrasoft pseudopotentials. In
reciprocal space the wave function was sampled at eight points
arranged in a Monkhorst-Pack grid \cite{Monkhorst}. The atomic
positions in the experimentally suggested model were varied until
the local energy minimum was found and the expected constant
current STM profile was obtained from the electronic structure
using the Tersoff-Hamann scheme \cite{Tersoff}.

Figure \ref{fig:sub2:b} shows the filled states image so obtained
at a sample bias of $-$2.0\,V. The image dimensions are 4.2
$\times$ 3.3 nm$^{2}$ and it can be directly compared with the
experimental result in figure \ref{fig:sub2:a}. The dominance of
Ho in the nanorod is evident in the modelled system. There are
many filled electronic states around the high-valency holmium
atoms that have a favourable probability for tunneling into the
tip. The trenchlike structure between the nanorods does indeed
seem to be formed by the arrangement of Ge that we have suggested.

Figure \ref{fig:sub3:b} shows the empty states image calculated
with a sample bias of +1.5\,V. The image dimensions are 4.2
$\times$ 3.3 nm$^{2}$ and it can be directly compared with the
experimental result in figure \ref{fig:sub3:a}. Atomic resolution
within the nanorod is apparent in the theoretical image as it was
in the experiment. Between the nanorods we see the structure in
the germanium underlayers that was not accessible in the
experiment.

\begin{figure}[h]
\centering \subfigure[] {
    \label{fig:sub5:a}
    \includegraphics[scale=0.4]{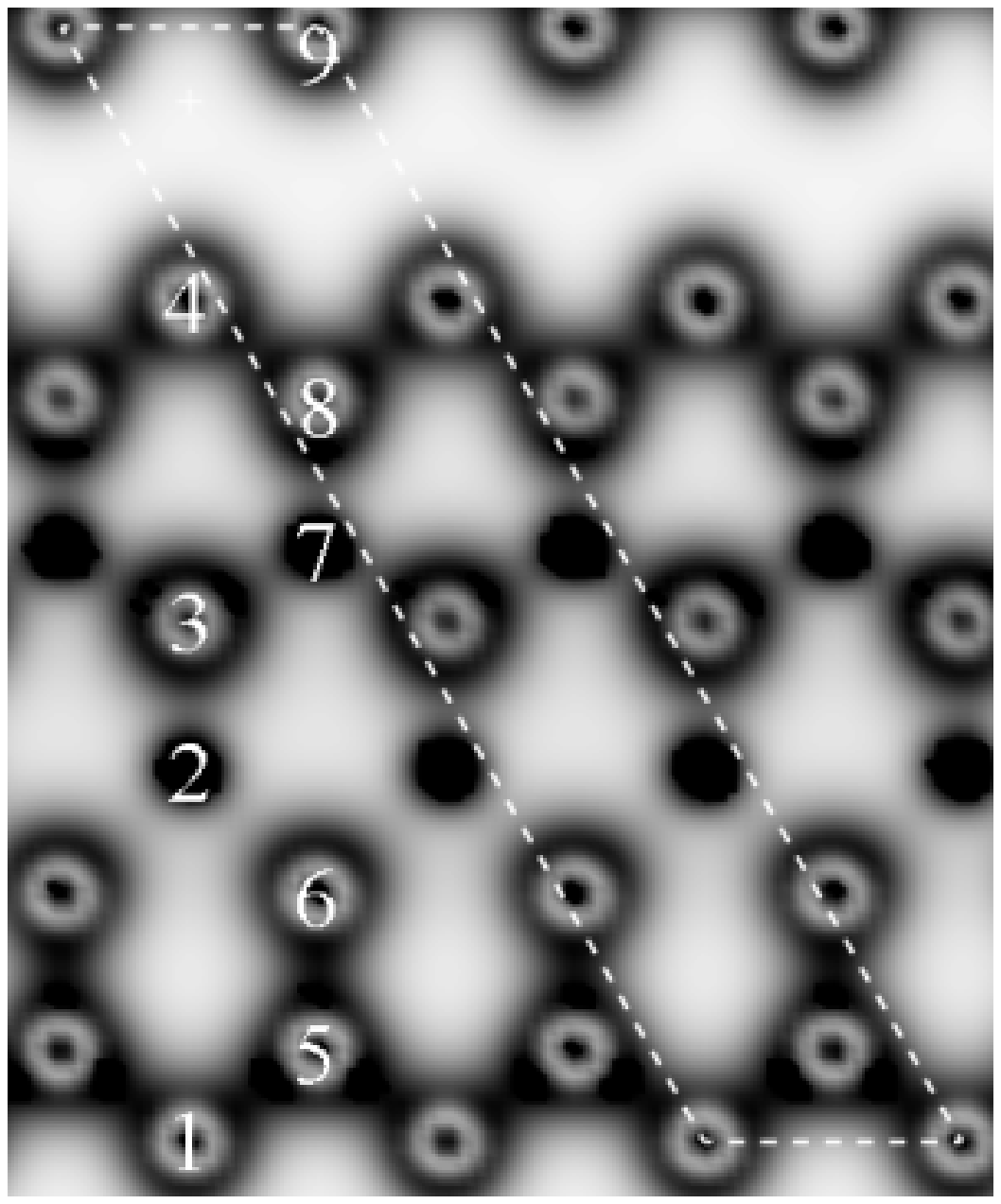}
} \hspace{1cm} \subfigure[] % caption for subfigure b
{
    \label{fig:sub5:b}
    \includegraphics[scale=0.4]{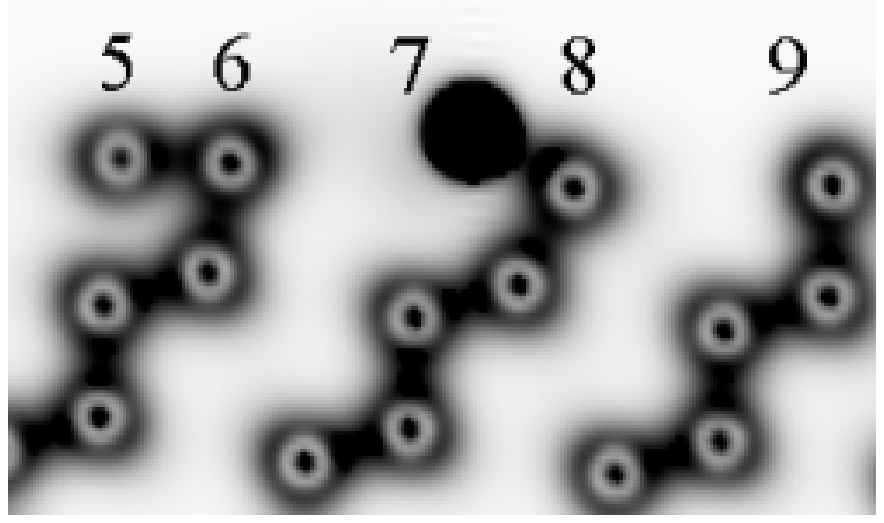}
} \hspace{1cm} \subfigure[] % caption for subfigure b
{
    \label{fig:sub5:c}
    \includegraphics[scale=0.4]{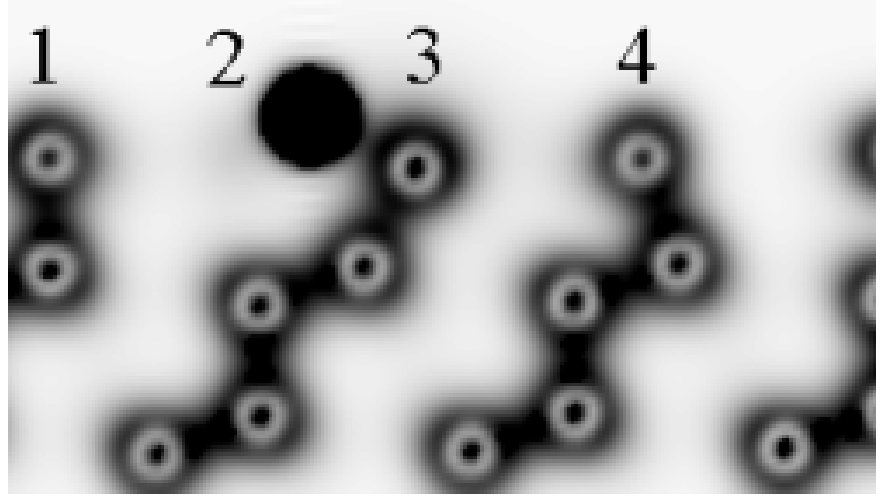}
}

\caption{Electron density within a slice taken (a) horizontally in
the plane of the surface through the flat layer of Ge atoms
(b)/(c) vertically through the two Ho atoms (labelled 2 and 7) to
show the bonding to the layer below.}
\label{fig:sub5} % caption for the whole figure
\end{figure}

The electronic structure can be used in conjunction with
population analysis to determine the bonding environment
responsible for the nanorod structure. Figure \ref{fig:sub5} shows
three slices through the electronic density, one in the plane of
the Ge flat layer and two vertical slices through the two holmium
atoms. There is a mixture of covalent and ionic bonding within the
nanorod reflecting the large contribution of electron transfer
from the Ho atoms.

In figure \ref{fig:sub5} Ge atom 3 is covalently bonded to the two
Ho atoms and to the Ge atom in the layer below in a tetrahedral
arrangement and this atom has negligible extra charge transferred
from either Ho atom. There is a significant amount of charge
transfer from Ho 2 to germanium 6 and to a lesser extent Ge 5 and
Ge 1 near to the edge of the shelf. Ge 5 is sp$^{2}$ hybridised
and we can see three planar trigonal bonds in the overhead view
and no bonds to the layer below in the side view. There is also a
significant amount of charge transfer from Ho 7 to Ge 8 and to a
lesser extent to Ge 4. Ge 1 and Ge 4 are partially sp$^{2}$
hybridised with some remainder of tetrahedral bonding. Charge
transfer from the two Ho atoms seems to be a key element in the
stability of this system.

\begin{figure}[h]
\flushleft
  \includegraphics[scale=1.0]{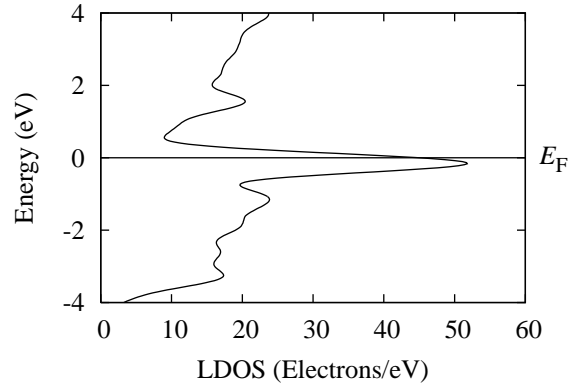}
  \caption{The calculated total local density of states for the
nanorod structure. There is a large concentration of states close
to the Fermi level that are thermally accessible that render the
system metallic.}\label{fig6}
\end{figure}

The electronic properties of the nanorods can be predicted from
the results of the \emph{ab initio} calculation. In figure
\ref{fig6} the local density of states within the nanorod is
shown. The spike in the electronic population at energies close to
the Fermi level indicates the metallic character of the nanorod
and confirms the decoupling of its electronic states from those in
the semiconducting germanium bulk.

\begin{figure}[h]
\flushleft
  \includegraphics{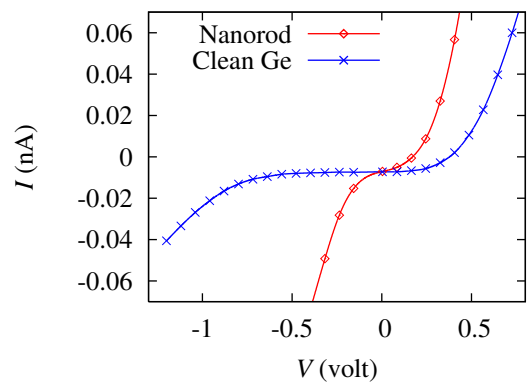}
  \caption{(Colour online) Experimental STS data showing the conducting properties of the
  nanorod. Data taken from the Ge substrate, with its large band gap at
  the Fermi level, is included for reference.}\label{fig7}
\end{figure}

Scanning Tunneling Spectroscopy data taken from the surface
supports this. Figure \ref{fig7} shows tunneling current
measurements taken from the nanorod and from the surrounding Ge
substrate for reference. The nanorod clearly has conducting states
at the Fermi level whereas the band gap of the Ge substrate means
that it has no conducting states at the Fermi level.

In conclusion, nanorods have been formed by depositing a low
coverage of Ho on the Ge(111) surface. These are a true isolated
nanostructure because they are not part of a periodic
reconstruction or rectangular islands. The nanorods have constant
width that is very narrow compared to other nanorod structures and
they do not require step edges or patterning in order to form.
When the experiment is repeated using a higher Ho coverage the
nanorods exist as part of a periodic 5$\times1$ structure. We have
introduced a model for this structure which we have quantitatively
validated using an \textit{ab initio} geometry optimisation. We
have shown that in both filled and empty states imaging the
calculated STM profile for this model is in good qualitative
agreement with experiment and the calculated electronic structure
suggests that the nanorod is metallic in character and can be
termed a nanowire.

\end{document}